\documentclass[twoside,twocolumn,9pt]{article}
\usepackage{extsizes}
\usepackage[super,sort&compress,comma]{natbib} 
\usepackage[version=3]{mhchem}
\usepackage[left=1.5cm, right=1.5cm, top=1.785cm, bottom=2.0cm]{geometry}
\usepackage{balance}
\usepackage{times,mathptmx}
\usepackage{sectsty}
\usepackage{graphicx} 
\usepackage{lastpage}
\usepackage[format=plain,justification=justified,singlelinecheck=false,font={stretch=1.125,small,sf},labelfont=bf,labelsep=space]{caption}
\usepackage{float}
\usepackage{fancyhdr}
\usepackage{fnpos}
\usepackage[english]{babel}
\addto{\captionsenglish}{}
\usepackage{array}
\usepackage{droidsans}
\usepackage{charter}
\usepackage[T1]{fontenc}
\usepackage[usenames,dvipsnames]{xcolor}
\usepackage{setspace}
\usepackage[compact]{titlesec}
\usepackage{hyperref}

\definecolor{cream}{RGB}{222,217,201}

\usepackage{soul}
\usepackage[normalem]{ulem}
\newcommand{\editor}[2]{%
  \expandafter\newcommand\csname #1note\endcsname[1]{%
    \textcolor{#2}{(\textbf{#1:} ##1)}}%
  \expandafter\newcommand\csname #1\endcsname[1]{%
    \textcolor{#2}{##1}}%
  \expandafter\newcommand\csname #1cancel\endcsname[1]{%
    \textcolor{#2}{\sout{##1}}}%
  \expandafter\newcommand\csname #1change\endcsname[2]{%
    \textcolor{#2}{\sout{##1} ##2}}%
  \newenvironment{#1text}{\color{#2}}{\color{black}}
}
\editor{IT}{red}
\editor{MC}{magenta}

\begin{document}

\pagestyle{fancy}
\thispagestyle{plain}
\fancypagestyle{plain}{

\renewcommand{\headrulewidth}{0pt}
}

\makeFNbottom
\makeatletter
\renewcommand\LARGE{\@setfontsize\LARGE{15pt}{17}}
\renewcommand\Large{\@setfontsize\Large{12pt}{14}}
\renewcommand\large{\@setfontsize\large{10pt}{12}}
\renewcommand\footnotesize{\@setfontsize\footnotesize{7pt}{10}}
\makeatother

\renewcommand{\thefootnote}{\fnsymbol{footnote}}
\renewcommand\footnoterule{\vspace*{1pt}%
\color{cream}\hrule width 3.5in height 0.4pt \color{black}\vspace*{5pt}} 
\setcounter{secnumdepth}{5}

\makeatletter 
\renewcommand\@biblabel[1]{#1}            
\renewcommand\@makefntext[1]%
{\noindent\makebox[0pt][r]{\@thefnmark\,}#1}
\makeatother 
\renewcommand{\figurename}{\small{Fig.}~}
\sectionfont{\sffamily\Large}
\subsectionfont{\normalsize}
\subsubsectionfont{\bf}
\setstretch{1.125} 
\setlength{\skip\footins}{0.8cm}
\setlength{\footnotesep}{0.25cm}
\setlength{\jot}{10pt}
\titlespacing*{\section}{0pt}{4pt}{4pt}
\titlespacing*{\subsection}{0pt}{15pt}{1pt}

\fancyfoot{}
\fancyfoot[LO,RE]{\vspace{-7.1pt}\includegraphics[height=9pt]{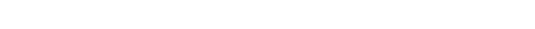}}
\fancyfoot[CO]{\vspace{-7.1pt}\hspace{13.2cm}\includegraphics{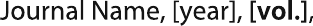}}
\fancyfoot[CE]{\vspace{-7.2pt}\hspace{-14.2cm}\includegraphics{head_foot/RF}}
\fancyfoot[RO]{\footnotesize{\sffamily{1--\pageref{LastPage} ~\textbar  \hspace{2pt}\thepage}}}
\fancyfoot[LE]{\footnotesize{\sffamily{\thepage~\textbar\hspace{3.45cm} 1--\pageref{LastPage}}}}
\fancyhead{}
\renewcommand{\headrulewidth}{0pt} 
\renewcommand{\footrulewidth}{0pt}
\setlength{\arrayrulewidth}{1pt}
\setlength{\columnsep}{6.5mm}
\setlength\bibsep{1pt}

\makeatletter 
\newlength{\figrulesep} 
\setlength{\figrulesep}{0.5\textfloatsep} 

\newcommand{\topfigrule}{\vspace*{-1pt}%
\noindent{\color{cream}\rule[-\figrulesep]{\columnwidth}{1.5pt}} }

\newcommand{\botfigrule}{\vspace*{-2pt}%
\noindent{\color{cream}\rule[\figrulesep]{\columnwidth}{1.5pt}} }

\newcommand{\dblfigrule}{\vspace*{-1pt}%
\noindent{\color{cream}\rule[-\figrulesep]{\textwidth}{1.5pt}} }

\makeatother

\twocolumn[
\begin{@twocolumnfalse}
{\includegraphics[height=30pt]{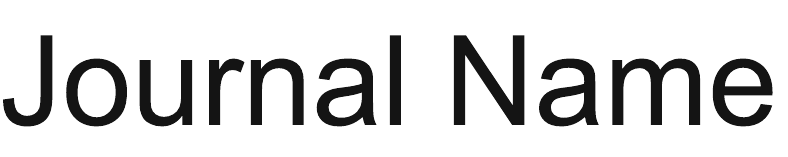}\hfill%
 \raisebox{0pt}[0pt][0pt]{\includegraphics[height=55pt]{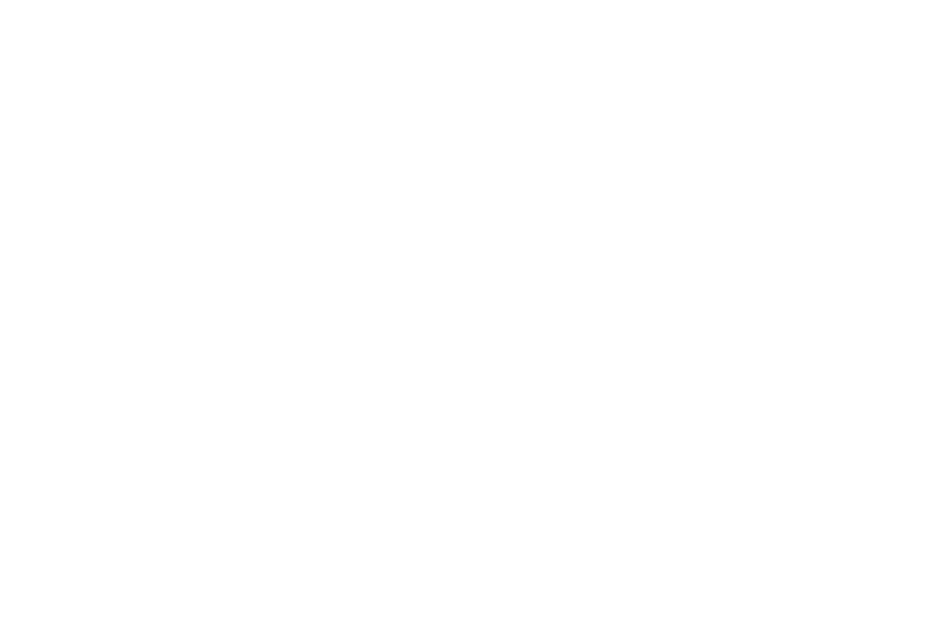}}%
 \\[1ex]%
 \includegraphics[width=18.5cm]{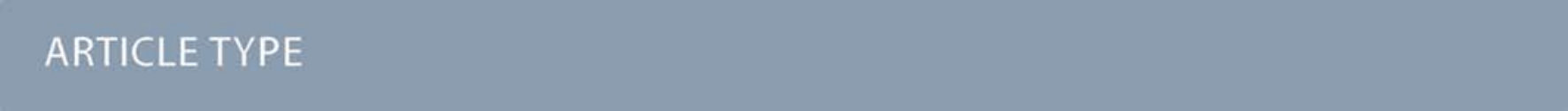}}\par
\vspace{3\parsep}
\sffamily
\begin{tabular}{m{4.5cm} p{13.5cm}}

\includegraphics{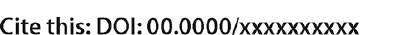} & \noindent\LARGE{\textbf{Optimizing accuracy and efficacy in data-driven materials discovery for the solar production of hydrogen}} \\   
\vspace{0.3cm} & \vspace{0.3cm} \\

& \noindent\large{Y.~Xiong,$^{a,\star}$ Q.~T.~Campbell,$^{b,\star}$ J.~Fanghanel,$^{a,c}$ C.~K.~Badding,$^{d}$ H.~Wang,$^{a}$ N.~E.~Kirchner-Hall,$^{a}$ M.~J.~Theibault,$^{d}$ I.~Timrov,$^{e}$ J.~S.~Mondschein,$^{c}$ K.~Seth,$^{c}$ R.~Katz,$^{c}$ A.~Molina Villarino,$^{d}$ B.~Pamuk,$^{f}$ M.~E.~Penrod,$^{a}$ M.~M.~Khan,$^{a}$ T.~Rivera,$^{c}$ N.~C.~Smith,$^{g}$ X.~Quintana,$^{a}$ P.~Orbe,$^{a}$ C.~J.~Fennie,$^{f}$ S.~Asem-Hiablie,$^{h}$ J.~L.~Young,$^i$ T.~G.~Deutsch,$^i$ M.~Cococcioni,$^{j}$ V.~Gopalan,$^{a}$ H.~D.~Abru\~na,$^{d}$ R.~E.~Schaak,$^{c}$ I.~Dabo$^{a,h}$} \\

\includegraphics{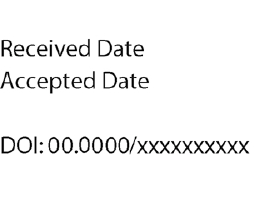} & \noindent\normalsize{The production of hydrogen fuels, \textit{via} water splitting, is of practical relevance for meeting global energy needs and mitigating the environmental consequences of fossil-fuel-based transportation. Water photoelectrolysis has been proposed as a viable approach for generating hydrogen, provided that stable and inexpensive photocatalysts with conversion efficiencies over 10\% can be discovered, synthesized at scale, and successfully deployed (Pinaud \textit{et al.}, \textit{Energy Environ.~Sci.}, 2013, \textbf{6}, 1983).} While a number of first-principles studies have focused on the data-driven discovery of photocatalysts, in the absence of systematic experimental validation, the success rate of these predictions may be limited. We address this problem by developing a screening procedure with co-validation between experiment and theory to expedite the synthesis, characterization, and testing of the computationally predicted, most desirable materials. Starting with 70,150 compounds in the Materials Project database, the proposed protocol yielded 71 candidate photocatalysts, 11 of which were synthesized as single-phase materials. Experiments confirmed hydrogen generation and favorable band alignment for 6 of the 11 compounds, with the most promising ones belonging to the families of alkali and alkaline-earth indates and orthoplumbates. This study shows the accuracy of a nonempirical, Hubbard-corrected density-functional theory method to predict band gaps and band offsets at a fraction of the computational cost of hybrid functionals, and outlines an effective strategy to identify photocatalysts for solar hydrogen generation.

\end{tabular}
\end{@twocolumnfalse} \vspace{0.6cm}
]

\renewcommand*\rmdefault{bch}\normalfont\upshape
\setcounter{secnumdepth}{0} 
\rmfamily
\section*{}
\vspace{-1cm}

\footnotetext{\textit{$^{a}$~Department of Materials Science and Engineering, and Materials Research Institute, The Pennsylvania State University, University Park, PA, USA}}
\footnotetext{\textit{$^{b}$~Sandia National Laboratories, Albuquerque, NM, USA}}
\footnotetext{\textit{$^{c}$~Department of Chemistry and Materials Research Institute, The Pennsylvania State University, University Park, PA, USA}}
\footnotetext{\textit{$^{d}$~Department of Chemistry and Chemical Biology, Cornell University, Ithaca, NY, USA}}
\footnotetext{\textit{$^{e}$~Theory and Simulation of Materials (THEOS) and National Centre for Computational Design and Discovery of Novel Materials (MARVEL), \'Ecole Polytechnique F\'ed\'erale de Lausanne, CH-1015 Lausanne, Switzerland}}
\footnotetext{\textit{$^{f}$~School of Applied and Engineering Physics, Cornell University, Ithaca, NY, USA}}
\footnotetext{\textit{$^{g}$~Department of Materials Science and Engineering, Northwestern University, Evanston, IL, USA}}
\footnotetext{\textit{$^{h}$~Institutes of Energy and the Environment, The Pennsylvania State University, University Park, PA, USA}}
\footnotetext{\textit{$^{i}$~National Renewable Energy Laboratory, Golden, CO, USA}}
\footnotetext{\textit{$^{j}$~Department of Physics, University of Pavia, Pavia, Italy}}

\footnotetext{\textit{$^\star$These authors contributed equally.}}
\footnotetext{$^\dag$~Electronic Supplementary Information (ESI) available: criteria of the literature survey, computed band gaps and band edges, optical and electrochemical measurements, synthesis references, and stability analysis. See DOI: 00.0000/00000000.}


\pagebreak

\section{\label{sec:intro}Introduction}

\begin{figure}
	\centering
	\includegraphics[width=\columnwidth]{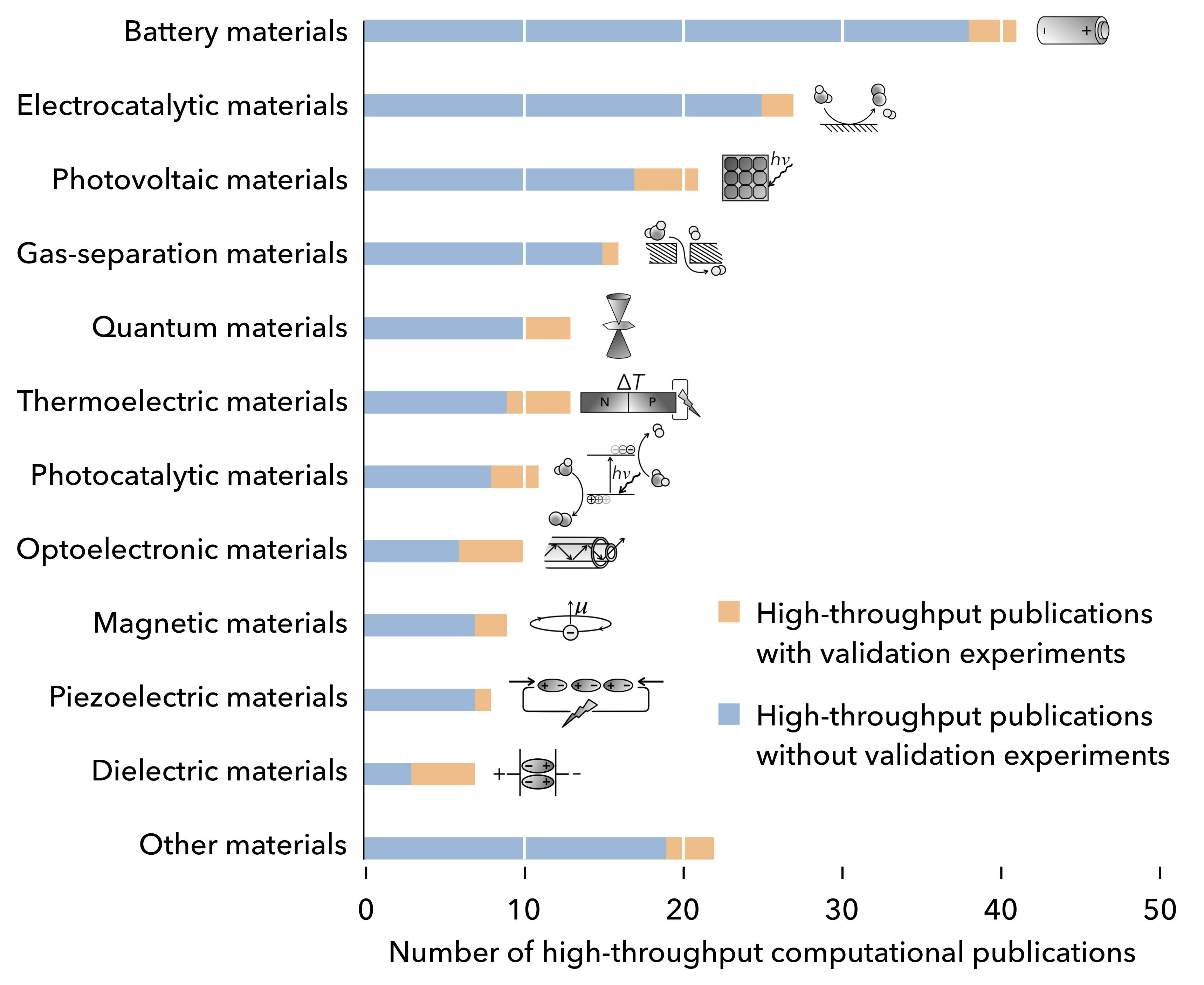}
	\caption{Survey of peer-reviewed publications in high-throughput computational materials science (source:~Web of Science; period:~2000-2020), organized into technological areas with the number of articles containing experimental validation indicated in orange and not containing experimental validation indicated in blue. Although not exhaustive, this survey is representative of the proportion (on the order of 20\%) of high-throughput computational predictions that are accompanied with validation experiments. (The criteria of this survey are explained in Fig.~S1, ESI$^\dag$.)}
	\label{figure:survey}
\end{figure}

\begin{figure*}
	\centering
	\includegraphics[width=\textwidth]{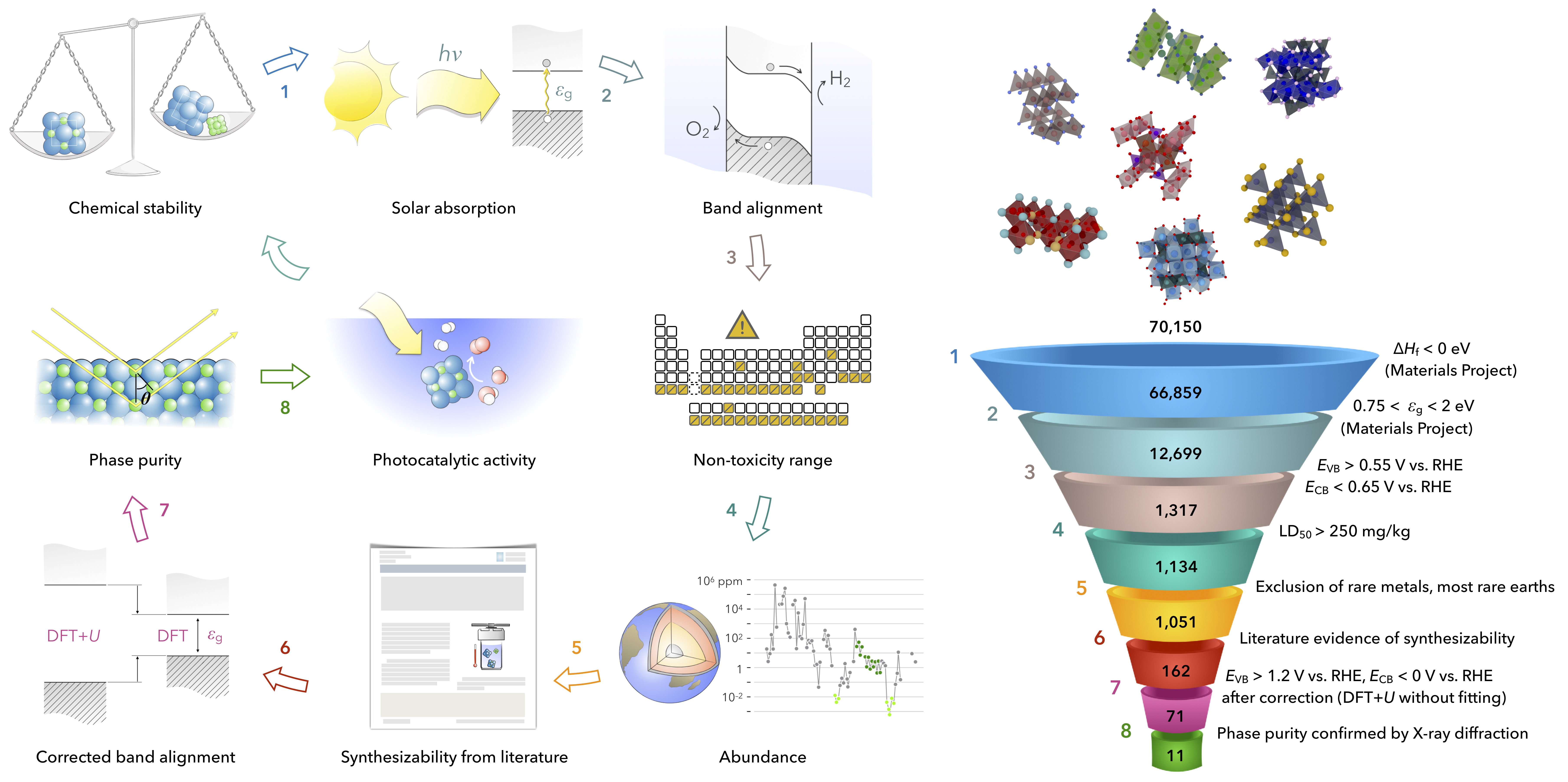}
	\caption{Screening protocol with co-validation between experiment and theory to identify photocatalytic semiconductors for hydrogen generation (left panel). The criteria of selection are listed next to the corresponding tiers of the funnel (right panel). Starting with 70,150 materials, 11 potential photocatalysts are selected by calculating their chemical stability, bang gap, and band alignment, and by probing their phase purity. The Hubbard \textit U parameters are determined without empirical fitting using linear-response theory. The enthalpy of formation, the band gap, the valence band edge, and the conduction band edge are denoted as $\Delta H_{\rm f}$, $\epsilon_{\rm g}$, $E_{\rm VB}$, and $E_{\rm CB}$, respectively; LD\textsubscript{50} stands for the median lethal dose of the constituent elements.}
	\label{figure:protocol}
\end{figure*}

Hybrid and plug-in electric vehicles have helped curtail the global consumption of petroleum-based fuels for personal transportation.\cite{Cazzola:2017,Melaina:2016} Yet the development of electric power systems satisfying the performance requirements of freight transport and air travel has faced major technical hurdles.\cite{Sarlioglu:2015,Sripad:2018} The global demand for transportation fuels has thus continued to increase at a rate of 1.2\% per year,~\cite{AnnualEnergyOutlook:2020} prompting efforts to advance renewable fuels such as hydrogen.\cite{Tachibana:2012,Montoya:2017} While hydrogen can be converted into electricity without carbon emissions, conventional methods to obtain hydrogen mostly involve steam reforming, a process that releases carbon dioxide.\cite{Spath:2000} Hence, there is a compelling need for a carbon-neutral supply of hydrogen, not only to develop sustainable transportation\cite{Offer:2010} but also to minimize greenhouse gas emissions from various industries such as the production of ammonia fertilizers, which requires hydrogen feedstocks.\cite{Sukalac:2009} 

Photocatalysis offers a potential solution for the solar generation of hydrogen by electrochemically cleaving water.\cite{Takata:2019} Feasibility analyses have shown that photoelectrolysis could be economically and technically viable to produce hydrogen industrially, with costs ranging from \$1.6 to \$3.2 per gasoline gallon equivalent, depending on the mode of photogeneration.\cite{Pinaud:2013} Despite these prospects, the photoelectrochemical production of hydrogen has been hindered by the lack of stable and inexpensive materials with solar-to-hydrogen conversion efficiencies exceeding the estimated threshold of 10\% for cost competitiveness.\cite{Pinaud:2013} 

Data-driven materials screening could expedite the discovery and development of efficient photocatalysts. \cite{Montoya2017,Setyawan2010, Calderon2015,Sohier2018,Armiento2011,Liu2013,Burton2018,Jain2011a} As depicted in Fig.~\ref{figure:survey}, high-throughput computational methods have been used to explore extensive databases of crystal structures in search for technological materials.\cite{Castelli:2012a,Castelli:2012b,Castelli:2015,Yan:2017,dePablo:2019} Although Fig.~\ref{figure:survey} shows encouraging results in using first-principles calculations to identify candidate compounds in areas as diverse as electrochemistry, photovoltaics, optoelectronics, thermoelectrics, and piezoelectrics, only a fraction of these studies predicted materials that were experimentally confirmed---with the notable exception of Ref.~\citenum{Yan:2017}, which focused on the computational and experimental discovery of vanadate-based oxides for photocatalytic oxygen evolution. This outcome is mainly due to the limited precision of conventional first-principles simulations, which rely on simplified descriptions of electronic interactions, and to existing collaborative barriers between experiment and theory. The goal of this work is to maximize the efficacy and success rate of first-principles methods for the data-driven discovery of water-splitting photocatalysts by providing a systematic experimental assessment of their predictive accuracy.

\section{\label{section:results}Results and discussion}

\subsection{\label{section:screening}Screening and synthesis}

The high-throughput screening procedure is depicted in Fig.~\ref{figure:protocol}. Starting from the Materials Project database,\cite{Jain:2013} we carried out an initial parsing of 70,150 compounds by retaining those whose enthalpy of formation $\Delta H_{\rm f}$ is predicted to be negative with respect to the reference states of their elements. We then identified the materials whose computed band gap $\epsilon_{\rm g}$ was between 0.75 eV and 2 eV. While the solar-to-hydrogen conversion efficiency of photocatalytic cells is maximal for $\epsilon_{\rm g}$ in the range of 1.5 to 2.5 eV,\cite{Hanna:2006} we rescaled this spectral window to account for the tendency of conventional density-functional theory (DFT)~\cite{Hohenberg:1964, Kohn:1965} approximations to underestimate the band gaps of semiconductors by a typical margin of 20-50\%,\cite{Perdew:1985} as detailed in Sec.~S2, ESI$\dag$. Applying these criteria to enthalpies and band gaps from the Materials Project, we narrowed the list down to 12,699 materials. 

Next, we inspected the valence band edge $E_{\rm VB}$ and conduction band edge $E_{\rm CB}$ of the candidate semiconductors relative to the redox potentials of the H\textsubscript2O/O\textsubscript2 and H\textsubscript2/H$^+$ couples, respectively. Although valence and conduction band edges can be predicted from first principles,~\cite{hormann2018} these predictions require supercell slab calculations that are computationally demanding and must be repeated for a representative set of surface facets, terminations, and adsorbates. Additionally, it is unclear how the band edges calculated for specific surfaces would relate to those of a polycrystalline material. In contrast, estimating the electrode potential from the electronegativities of the constituent elements has been shown to be reasonably accurate in predicting band alignments,~\cite{Butler1978} as will be further assessed and discussed in the next section. 

In explicit terms, band edges were calculated from the band gap $\epsilon_{\rm g}$ and geometric mean $\langle \chi \rangle$ of the Mulliken electronegativities of the constituent elements (which provides an estimate for the opposite Fermi energy $\epsilon_{\rm F}$ and flatband potential $E_{\rm FB}$ through $ E_{\rm FB} = -\epsilon_{\rm F}/e = \langle \chi \rangle/e$, where $e$ is the fundamental charge of an electron). In the absence of Fermi-level pinning (and if the electron and hole effective masses are comparable in magnitude), one can evaluate the band edges as $E_{\rm VB} = (\langle \chi \rangle  + \epsilon_{\rm g} / 2)/e$ and $E_{\rm CB} = ( \langle \chi \rangle  - \epsilon_{\rm g} / 2 )/e$.\cite{Wu:2013} These calculations identified 1,317 candidates  fulfilling the conditions $E_{\rm VB}$ > 0.575 V and $E_{\rm CB}$ < 0.625 V on the reversible hydrogen electrode (RHE) scale (cf.~Fig.~S2, ESI$\dag$ for the derivation of these band-edge criteria). The list was further pruned by examining the toxicity of the individual elements on the LD\textsubscript{50} (median lethal dose) scale, and by taking into account their crustal abundance and radioactivity.  In specific terms, we eliminated elements with an LD\textsubscript{50} value lower than 250 mg/kg\cite{IDLH:2014} and those that are labeled as radioactive in the Evaluated Nuclear Structure and Decay database of the International Atomic Energy Agency.\cite{Bhat:1992} Rare-earth elements and transition metals with an abundance lower than that of gold (0.0004 ppm by mass) were also removed, yielding 1,051 candidates.\cite{Yaroshevsky:2006}

In addition, we identified materials that were amenable to experimental synthesis. To this end, we referenced the 1,051 compounds to the Crystallography Open Database,~\cite{Gravzulis:2009} finding 452 materials that had been previously synthesized. A systematic search of the experimental literature enabled us to establish a list of materials potentially accessible through conventional synthesis techniques based on criteria of availability of the chemical precursors, sintering time, and likelihood of phase purity. We also incorporated high-level insights pertaining to reactivity and stability; for example, we eliminated compounds that are likely, based on their expected chemical reactivity, to be sensitive to air or water. This analysis ultimately led to a database of 162 materials whose synthesis steps are outlined in the Synthesizability section, ESI$^\dag$. In parallel, we evaluated the performance of machine-learning algorithms in suggesting synthesis actions and chemical precursors.\cite{Kim:2017a,Kim:2017b, Jensen:2019} This preliminary evaluation indicates that machine learning may ultimately be used for assessing materials synthesizability. As machine-learning models continue to be developed, we will explore the possibility of incorporating these capabilities into data-driven screening protocols.

From this list, we refined the electronic-structure predictions of the band gaps. Although the semilocal DFT approximation\cite{Perdew:1996} is generally apt at predicting the stability and reactivity of covalent materials, it is known to underestimate their band gap $\epsilon_{\rm g}$ by a considerable margin.\cite{Cohen:2008} The limitations of conventional functionals for determining band gaps originate from self-interaction errors~\cite{Perdew:1981, MoriSanchez:2006} that cause electrons to delocalize and to thus become unphysically prone to optical excitation. Many-body perturbation theories~\cite{Hedin1965,Onida2002} and hybrid functionals~\cite{Heyd2003,Skone2014,Skone2016} are among the most accurate electronic-structure methods to predict the band gaps of semiconductors. These methods improve over conventional DFT approximations by capturing the spectral or nonlocal dependence of the single-electron potential,~\cite{Ferretti:2014} which may lead, however, to a significant increase in algorithmic complexity for unit cells with a few hundreds of valence electrons (typically, 50-100 atoms). 

A widely used approach to determine band gaps while preserving computational efficiency consists of incorporating localization terms at atomic sites (the Hubbard \textit{U} method),\cite{Anisimov1991, Anisimov1997, Dudarev1998, Kulik:2006, Kulik:2008} with the caveat that the magnitude of these terms is in general not known \textit{a priori}. Thus, the Hubbard \textit{U} parameters are frequently fitted to experimental band structures and thermodynamic energies. This approach yields \textit{U} parameters that may not be transferable from one compound (or crystal phase) to another, and it may not guarantee that other properties (such as lattice parameters and magnetization) are in better agreement with experiments. In addition, this fitting strategy is not applicable to materials for which limited experimental data are available, precluding its use within first-principles workflows. To overcome these limitations, we exploit a newly developed computational procedure,~\cite{Timrov2018} which enables us to determine the on-site \textit{U} parameters from linear-response theory~\cite{Cococcioni2005} with high efficiency and without relying on empirical fitting, as detailed in the Methods section. In this approach,~\cite{Timrov2018} density-functional perturbation theory (DFPT) is used to determine the on-site Hubbard parameters from the response of the system to a series of wavevector-modulated atomic potential shifts. This method offers several advantages compared to the previous implementation,~\cite{Cococcioni2005} among which high control on the numerical accuracy of the \textit U parameters and full automation of their calculations, making it suitable for high-throughput screening.

\begin{figure}[H]
    \centering
	\includegraphics[width=\columnwidth]{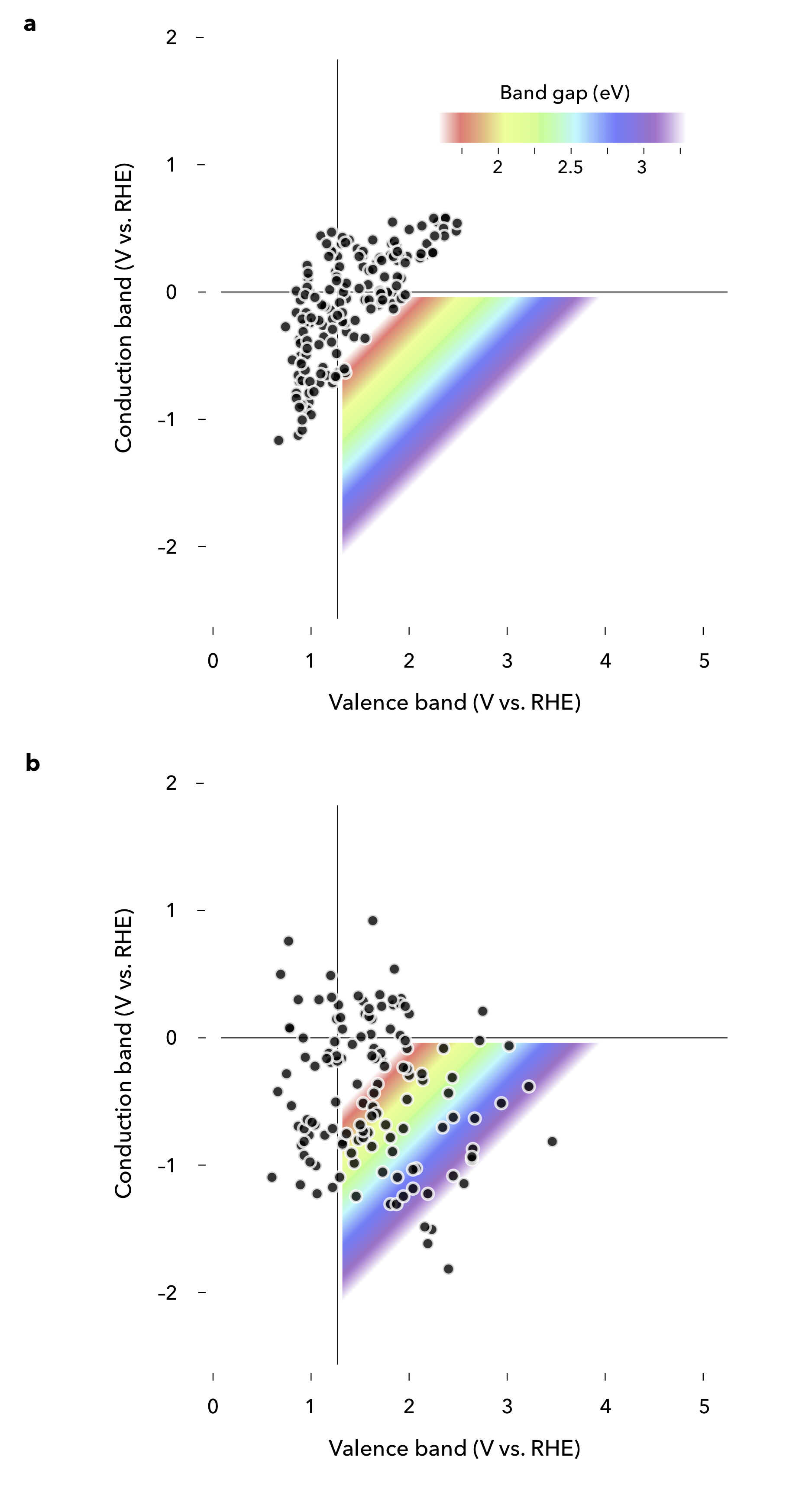}
	\caption{Band edges of the 162 candidate photocatalysts (obtained by applying criteria of chemical stability, band gap, band alignment, elemental toxicity, crustal abundance, and compound synthesizability), calculated (a) with fixed Hubbard \textit U parameters (DFT+\textit U\textsubscript{MP}) and (b) with Hubbard \textit U parameters determined from first principles (DFT+\textit U). In these diagrams, the rainbow region represents the visible optical range, where each colored, diagonal isoline corresponds to a constant band gap. Candidate photocatalytic materials occupy this domain.} \label{figure:rainbow}
\end{figure}

The impact of this correction on $\epsilon_{\rm g}$, $E_{\rm VB}$ and $E_{\rm CB}$ can be directly appreciated by comparing our \textit{ab initio} DFT+\textit U calculations with the predictions from the Materials Project (DFT+\textit U\textsubscript{MP}), which rely on constant \textit U parameters, optimally tuned to reproduce experimental enthalpies of formation (Table S2, ESI$^\dag$). Figure \ref{figure:rainbow} shows strong shifts in the distribution of the candidate materials, corresponding to a typical increase in the band gap of several eVs. It is seen that DFT+\textit{U} predictions idenfity numerous potential photocatalysts in the water-splitting region. A complete list of the band gaps and band edges for the 162 candidates is provided in Table S1 (DFT+\textit U) and Table S2 (DFT+\textit U\textsubscript{MP}), ESI$^\dag$.

\begin{figure*}
	\centering
	\includegraphics[width=\textwidth]{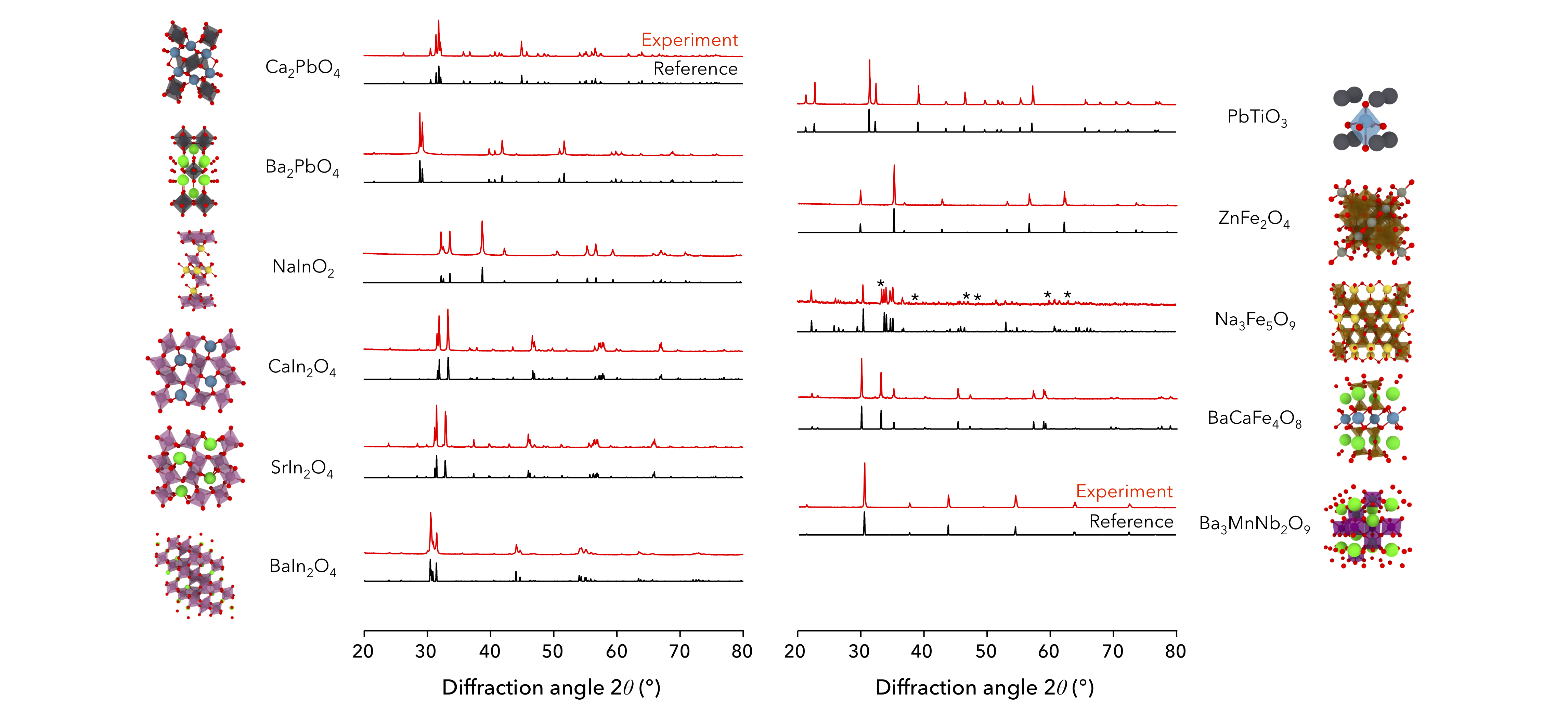}
	\caption{Comparison of the reference and measured X-ray diffraction patterns for the 11 compounds that were screened and synthesized. The peaks labeled with asterisks (*) in the Na\textsubscript3Fe\textsubscript5O\textsubscript9 spectrum are due to a secondary NaFe\textsubscript3O\textsubscript5 phase.
	\label{figure:diffraction}}
\end{figure*}

Based on these results, we restricted the list of potential photocatalysts to 71 compounds by imposing the criteria $E_{\rm VB}$ > 1.2 V vs.~RHE and $E_{\rm CB}$ < 0 V vs.~RHE. Powder samples of a subset of these compounds were synthesized \textit{via} solid-state reactions involving the mixing of precursors and calcining these at high temperatures for a given number of hours, as described in the Methods section. We focused on oxides, as these were able to be synthesized in a reasonable amount of time using standard furnace reactions. Figure~\ref{figure:diffraction} shows the normalized experimental X-ray diffraction (XRD) patterns, which closely match the reference patterns in most cases, confirming the synthesis of the expected single phase. We note that sodium ferrite Na\textsubscript3Fe\textsubscript5O\textsubscript9 contained a small amount of a related secondary phase NaFe\textsubscript3O\textsubscript5, as well as a higher background (\textit{i.e.}~a comparatively lower signal to noise ratio) than the others, suggesting that part of the sample was amorphous. Also, CaIn\textsubscript2O\textsubscript4 was not found to be the expected cubic ($Fd\bar 3m$) polymorph but instead to be orthorhombic ($Pnma$). We thus recalculated the band gap and edges of CaIn\textsubscript2O\textsubscript4 in the orthorhombic phase, finding moderate changes of 0.2 eV in the orbital energies, as reported in Table S1, ESI$^\dag$. We then proceeded to the characterization of these compounds. 

\begin{figure}
	\centering
	\includegraphics[width=\columnwidth]{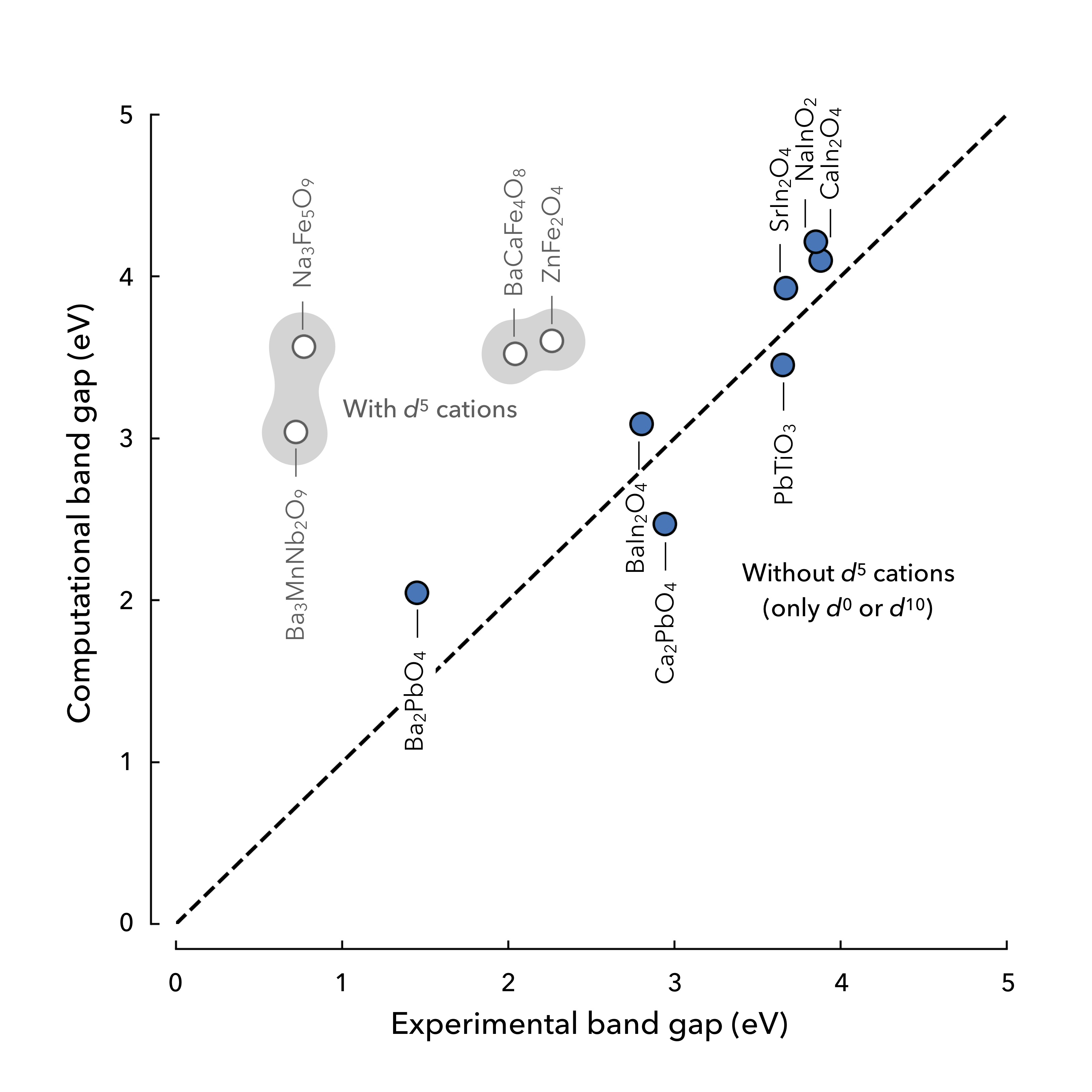}
	\caption{Comparison of the band gaps computed with the DFT+\textit{U} method to the experimental band gaps for the 11 single-phase compounds. The discrepancies observed for the 4 materials highlighted in grey can be ascribed to mid-gap defect levels and magnetic order due to open-shell \textit{d}\textsuperscript{5} transition metals. Compounds that contain \textit{d}\textsuperscript{5} cations are shown as white circles, while those that only contain \textit{d}\textsuperscript{0} and \textit{d}\textsuperscript{10} cations are shown as blue circles. Measurements of the band gaps are provided in Fig.~S4, ESI$\dag$, with the exception of PbTiO\textsubscript3 whose band gap is from Ref.\citenum{Wemple1970}.}
	\label{figure:bandgap}
\end{figure}

\begin{figure*}
	\centering
	\includegraphics[width=\textwidth]{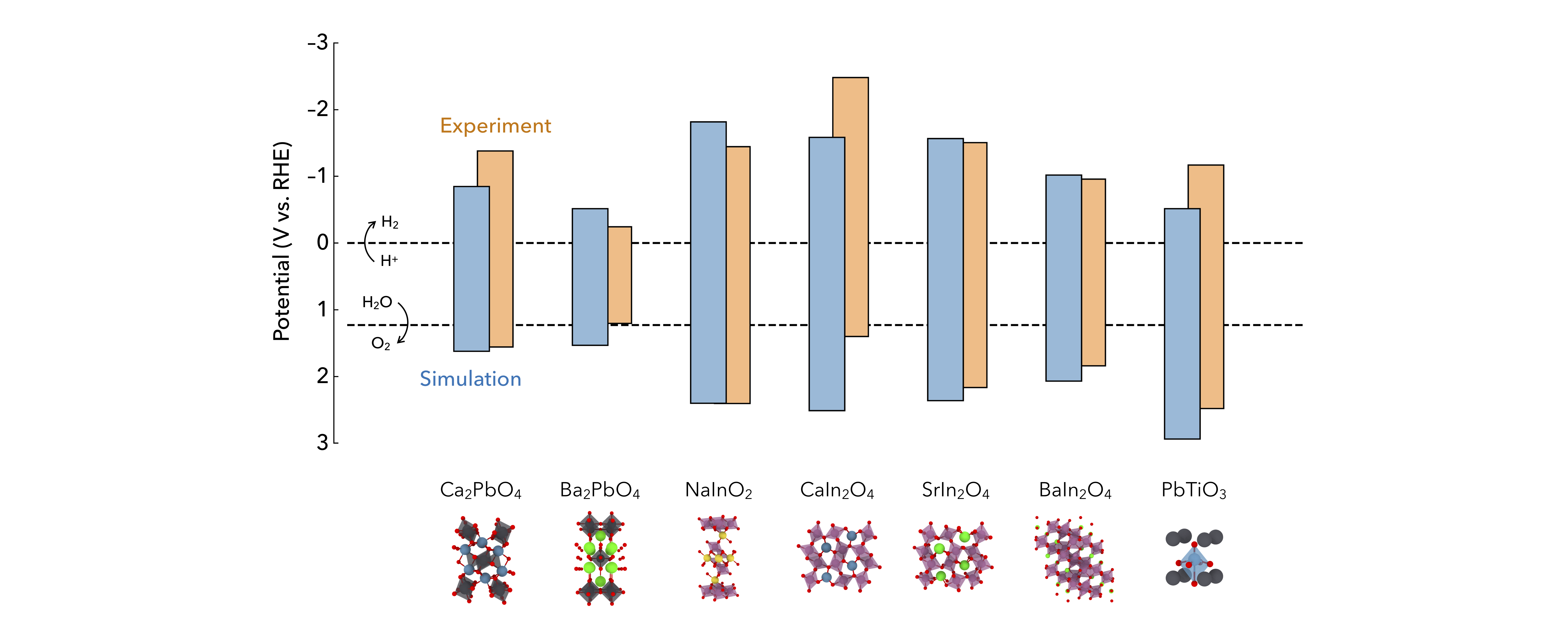}
	\caption{Experimentally determined band alignment (orange) with respect to the redox potentials of water (dashed lines), compared with the electronic energies estimated from the DFT+\textit{U} band gap for compounds with closed-shell (\textit{d}\textsuperscript{0} or \textit{d}\textsuperscript{10}) elements (cf.~Fig.~\ref{figure:bandgap}). Measurements of the band edges are shown in Fig.~S5, ESI$\dag$, with the exception of PbTiO\textsubscript3 whose band edges are from Ref.\citenum{Yang2019}.}
	\label{figure:bandedge}
\end{figure*}

\subsection{\label{section:characterization} Characterization and testing}

After synthesizing the candidate compounds, we compared their band gaps to computational predictions. As shown in Fig.~\ref{figure:bandgap}, this comparison highlights two distinct trends. For the materials that do not contain elements with partially filled (\textit{e.g.}~\textit{d}\textsuperscript5) shells (namely, the Fe and Mn oxides considered here), experimental and theoretical band gaps are found to be in close agreement, with a DFT+\textit{U} root mean squared error of 0.4 eV, which is considerably lower than the DFT error of 1.5 eV. The correspondence is quite remarkable for the alkaline-earth indates where the error does not exceed 0.3 eV. The largest DFT+\textit{U} deviation is found to be 0.55 eV for Ba\textsubscript2PbO\textsubscript4, whereas the DFT underestimation exceeds --0.9 eV for this material (cf.~Table S1, ESI$^\dag$). 

At variance with this predictive accuracy, the band gaps of the Fe and Mn oxides are significantly overestimated by DFT+\textit{U} calculations. These deviations possibly originate from an incorrect description of  magnetic order in these compounds. For example, a literature search reveals that zinc ferrite ZnFe\textsubscript2O\textsubscript4 adopts low-symmetry ferromagnetic order at room temperature\cite{Boucher:1970} with one of the spin channels exhibiting semi-metallicity,\cite{Soliman:2011} which cannot be captured straightforwardly by our collinear DFT+\textit U simulations. These magnetic characteristics complicate the accurate prediction of the band gap and the detailed interpretation of optical experiments---and is possibly at the origins of some observed inconsistencies. Yet the large discrepancy for Ba\textsubscript3MnNb\textsubscript2O\textsubscript9 and Na\textsubscript3Fe\textsubscript5O\textsubscript9 is more difficult to explain in terms of magnetism alone, and is instead likely due to the existence of mid-gap defects arising from the wide range of oxidation states that Mn and Fe can adopt. This interpretation is supported by the Tauc measurements, which show strong signatures of defect levels in the band gap, as it is apparent from the additional absorption peaks (the Kubelka--Munk function is proportional to absorption) within the main slope of the Tauc graphs reported in Fig.~S4, ESI$^\dag$.

To address this complexity, we have estimated the band gap as the extrapolation of the highest linear portion of the Tauc plot inside the region of prevalent decline in absorbance. This approach allows us to ascribe any absorption peak in this range to other energy states, such as trap or defect levels. This assignment makes intuitive sense as any mid-gap trap or defect state must have an energy difference (relative to the valence band) smaller than that of the energy gap. One of the reasons why these materials could exhibit a high density of trap states is that they contain transition metals with partially filled \textit{d} orbitals and variable oxidation states, which tend to produce energy levels within the band gap of a semiconductor. In fact, these effects are some of the known mechanisms whereby transition-metal-bearing materials acquire their colorful appearance. Although we are not discarding the creation of defect states caused by, \textit{e.g.}~vacancies and surface dangling bonds, these properties are seldom found in materials that do not contain open-shell transition metals. Accordingly, it is seen that there are far fewer disturbances in the Tauc plots of these compounds, which makes the analysis of the band gaps tractable and reliable within DFT+\textit U at a fraction of the computational cost of, \textit{e.g.}~hybrid functionals and many-body perturbation theories. A notable exception to these trends is PbTiO\textsubscript3, for which we did not find conclusive agreement between the Tauc measurement (2.69 eV) and available experimental data.~\cite{Wemple1970} This discrepancy is possibly due to the polar nature of PbTiO\textsubscript3, which affects the interpretation of the optical response.\cite{Wemple1970} We thus compared the DFT+\textit U band gap of PbTiO\textsubscript3 to the polarization-dependent band gaps of 3.27-3.38 eV from Ref.\citenum{Wemple1970}. Similarly, the flatband potential (--1.2 V vs. RHE) that we measured for PbTiO\textsubscript3 was unreasonably negative relative to previous experiments. Therefore, we compared the computed band edges to measurements from Ref.\citenum{Yang2019}, in accordance with the expected band gap ($\sim$3.3 eV).

Having determined the band gaps of the synthesized materials, we measured their band edges through Mott--Schottky plots to determine their flatband potentials. We monitored the reciprocal of the squared differential capacitance as a function of the applied potential and studied the linear region, following the Mott--Schottky equation (as exemplified for the case of an \textit{n}-type semiconductor):  ${1}/{C^2} = 2/(\epsilon_{\rm s} \epsilon_0 e N) \left((E - E_{\rm FB}) -({k_{\rm B}T})/e \right)$, where $C$ is the capacitance of the electrode, $\epsilon_{\rm s}$ is the dielectric constant of the semiconductor, $\epsilon_0$ is the vacuum permittivity, $N$ is the dopant concentration and $E-E_{\rm FB}$ is the applied potential relative to the flatband potential of the semiconductor. The Mott--Schottky plot of the majority of the compounds studied showed a single linear region, as shown in Fig.~S5, ESI$^\dag$. The horizontal intercept was used to determine the flatband potential. With the potential $E_{\rm FB}$ and the band gap $\epsilon_{\rm g}$ previously determined, we positioned the valence and conduction band edges on the redox scale in a manner identical to the computational evaluation: $E_{\rm CB} = E_{\rm FB} - \epsilon_{\rm g}/(2e)$ and $E_{\rm VB} = E_{\rm FB} + \epsilon_{\rm g}/(2e)$.

The computational results compiled in Fig.~\ref{figure:bandedge} and Table~\ref{table:compilation} are in qualitative agreement with experimental trends; for the majority of the compounds, computational predictions are within a few tenths of a volt from the measured redox potentials. In particular, the DFT+\textit{U} approach captures the anodic shifts in the conduction band edges as one moves down the alkaline-earth period for both the plumbate and indate series. Although these trends are correctly described, computational predictions appear to systematically overestimate the redox potentials, as it is clearly seen in the case of calcium orthoplumbate Ca\textsubscript2PbO\textsubscript4, calcium indate CaIn\textsubscript2O\textsubscript4, and lead titanate PbTiO\textsubscript3. In spite of these observations, all of the other calculated band edges are in close correspondence with experimental data, especially for SrIn\textsubscript2O\textsubscript4, where the predicted conduction and valence band edges are both within a mV from the measured potentials. Finally, we note that the largest deviation between theory and experiment is observed for CaIn\textsubscript2O\textsubscript4. This discrepancy may be indicative of negatively charged surface states, such as dangling bonds induced by oxygen vacancies, which can cause an increase in the Fermi energy (making it more negative on the redox scale). The possibility of surface states and their impacts on photocatalytic activity are examined below.

\begin{table*}[!htbp]
\centering
\caption{Band gap ($\epsilon_{\rm g}$), conduction and valence band edges ($E_{\rm CB}$ and $E_{\rm VB}$, respectively), flatband potential ($E_{\rm FB}$),  hydrogen production, and magnetic order (Y = yes, N = no) from electronic-structure calculations (DFT+\textit U), and from optical and electrochemical experiments. Hydrogen production is examined under two different test conditions: (\textit{i}) 0.1 M oxalic acid and (\textit{ii}) 15\% volume fraction of methanol in water. Hydrogen detection under conditions (\textit{i}) and (\textit{ii}) is indicated by a filled circle ($\bullet$) on the left and on the right, respectively. Similarly, the absence of hydrogen is denoted by an empty circle ($\circ$). The cross symbol ($\times$) indicates that the sample corrodes in the aqueous solution. Compounds whose electrochemical corrosion is not accompanied by a perceivable decrease in photocatalytic activity are indicated by a crossed, filled circle. The experimental band edges of the four Fe and Mn oxides could not be reliably determined due to mid-gap states, which are expected to set the position of the Fermi level.}
\label{table:compilation}
\begin{tabular}{l l c c c c r r r r r r r r r r}
\hline \\
{} & {} & Space & Magnetic & Hydrogen & \hspace{0.25cm} & \multicolumn{3}{c}{DFT+\textit{U} (eV)} & \hspace{0.25cm} & \multicolumn{4}{c}{Expt. (eV)} \\
{} & {} & group & order & production & & \multicolumn{1}{c}{$\epsilon_{\rm g}$} & \multicolumn{1}{c}{$eE_{\rm CB}$} & \multicolumn{1}{c}{$eE_{\rm VB}$} & & \multicolumn{1}{c}{$\epsilon_{\rm g}$} & \multicolumn{1}{c}{$eE_{\rm FB}$} & \multicolumn{1}{c}{$eE_{\rm CB}$} & \multicolumn{1}{c}{$eE_{\rm VB}$} \\
\\
\hline \\
1 & Ca\textsubscript2PbO\textsubscript4 & $Pbam$ & N & $\ooalign{$\circ$\cr$\hidewidth\times\hidewidth$\cr} \ooalign{$\bullet$\cr$\hidewidth\times\hidewidth$\cr}$ & & 2.47 & --0.79 & 1.68 & & 2.94 & 0.09 & --1.38 & 1.56 \\
2 & Ba\textsubscript2PbO\textsubscript4 & $I4/mmm$ & N & $\ooalign{$\circ$\cr$\hidewidth\times\hidewidth$\cr}  \ooalign{$\circ$\cr$\hidewidth\times\hidewidth$\cr} $ & & 2.05 & --0.46 & 1.59 & & 1.45 & 0.48 & --0.25 & 1.21  \\
3 & NaInO\textsubscript2 & $R \bar 3m$ & N & $\bullet \bullet$ & & 4.22 & --1.76 & 2.46 & & 3.85 & 0.48 & --1.45 & 2.41 \\
4 & CaIn\textsubscript2O\textsubscript4 & $Pnma$ & N & $ \circ \circ$ & & 4.10 & --1.58 & 2.52 & & 3.88 & --0.54 & --2.48 & 1.40 \\
5 & SrIn\textsubscript2O\textsubscript4 & $Pnma$ & N & $\bullet \circ $ & & 3.93 & --1.51 & 2.42 & & 3.67 & 0.33 & --1.51 & 2.17 \\
6 & BaIn\textsubscript2O\textsubscript4 & $P2_1/c$ & N & $\ooalign{$\circ$\cr$\hidewidth\times\hidewidth$\cr} \circ $ & & 3.09 & --0.96 & 2.13 & & 2.80 & 0.45 & --0.96 & 1.84  \\
7 & PbTiO\textsubscript3 & $P4mm$ & N & $\bullet \bullet $ & & 3.45 & --0.46 & 3.00 & & 3.3$^a$ & 0.5$^b$ & --1.2$^b$ & 2.2$^b$ \\
8 & ZnFe\textsubscript2O\textsubscript4 & $Fd\bar 3m$ & Y & $\bullet \bullet $ & & 3.60 & --0.38 & 3.22 & & 2.26 & 0.01 & --- & --- \\
9 & Na\textsubscript3Fe\textsubscript5O\textsubscript9 & $C2/c$ & Y & $\bullet \bullet $ & & 3.57 & --0.93 & 2.64 & & 0.77 & 0.23 & --- & ---- \\
10 & BaCaFe\textsubscript4O\textsubscript8 & $P\bar 31m$ & Y & $\ooalign{$\circ$\cr$\hidewidth\times\hidewidth$\cr}  \ooalign{$\circ$\cr$\hidewidth\times\hidewidth$\cr} $ & & 3.52 & --0.87 & 2.65 & & 2.04 & --0.44 & --- & --- \\
11 & Ba\textsubscript3MnNb\textsubscript2O\textsubscript9 & $P\bar 3m1$ & Y & $\circ \circ $ & & 3.04 & --0.70 & 2.34 & & 0.72 & --0.42 & --- & ---\\ \\
\hline
\end{tabular}
\\
\raggedright
\hspace{0.25cm} $^a$Ref.\citenum{Wemple1970}, $^b$Ref.\citenum{Yang2019}. 
\end{table*}

\begin{table}[!h]
\centering
\caption{Materials recommendations from the list of screened compounds. This list includes compounds with \textit{d}\textsuperscript{0} and \textit{d}\textsuperscript{10} transition-metal cations and main-group metal oxides which are not expected to induce mid-gap states and magnetic structures. The predicted band gaps are compared to experimental data, where available.}
\label{table:recommendations}
\begin{tabular}{l c r r r}
\hline \\
 & Space & \multicolumn{1}{c}{DFT} & \multicolumn{1}{c}{DFT+\textit{U}} & \multicolumn{1}{c}{Expt.} \\
 & group & \multicolumn{1}{c}{$\epsilon_{\rm g}$ (eV)} & \multicolumn{1}{c}{$\epsilon_{\rm g}$ (eV)} & \multicolumn{1}{c}{$\epsilon_{\rm g}$ (eV)} \\ \\
\hline \\
Oxides \\
\hspace{0.25cm} SrCu\textsubscript2O\textsubscript2 & $I4_1/amd$ & 1.81 & 3.11 & 3.3$^a$ \\
\hspace{0.25cm} BaCu\textsubscript2O\textsubscript2 & $I4_1/amd$ & 1.39 & 3.22 \\
\hspace{0.25cm} CuAlO\textsubscript2 & $R\bar 3m$ & 1.78 & 3.07 & 2.99$^b$ \\
\hspace{0.25cm} CuGaO\textsubscript2 & $R\bar 3m$ & 0.75 & 2.46 & 1.97$^b$ \\
\hspace{0.25cm} Ca\textsubscript{12}Al\textsubscript{14}O\textsubscript{33} & $C2$ & 2.01 & 3.73 & 4.17$^c$ \\
\hspace{0.25cm} Na\textsubscript3BiO\textsubscript4 & $P2/c$ & 1.04 & 2.21 \\ 
\hspace{0.25cm} Sr\textsubscript2PbO\textsubscript4 & $Pbam$ & 1.43 & 2.31 & 2.64$^d$\\
& & & & 1.75$^e$ \\ \\
Sulfides \\
\hspace{0.25cm} Cu\textsubscript3SbS\textsubscript3 & $P2_12_12_1$ & 1.06 & 1.89 & \\
\hspace{0.25cm} Cu\textsubscript2WS\textsubscript4 & $P\bar 42m$ & 1.66 & 2.06 & 1.74$^f$\\
\hspace{0.25cm} Cu\textsubscript3NbS\textsubscript4 & $P\bar 43m$ & 1.81 & 1.97 & 2.5$^g$\\ 
\hspace{0.25cm} CuYS\textsubscript2 & $Pnma$ & 1.63 & 2.18 & \\ \\
Oxychalcogenides \\
\hspace{0.25cm} LaOCuS & $P4/nmm $ & 1.70 & 2.65 & 3.1$^h$ \\
\hspace{0.25cm} LaOCuSe & $P4/nmm $ & 1.48 & 2.44 & 2.82$^h$ \\
\hspace{0.25cm} La\textsubscript4O\textsubscript4Se\textsubscript3 & $Amm2$ & 2.01 & 2.04 & 1.9$^i$ \\
\hspace{0.25cm} Na\textsubscript2TeO\textsubscript4 & $P2_1/c$ & 1.39 & 3.30 & \\ \\
Oxynitrides \\
\hspace{0.25cm} CaTaO\textsubscript2N & $Pmc2_1$ & 1.67 & 2.46 & 2.6$^j$\\ 
\hspace{0.25cm} LaTiO\textsubscript2N & $I2_1 2_1 2_1$ & 1.36 & 2.42 & 2.1$^k$\\ \\
Others \\
\hspace{0.25cm} Na\textsubscript5CuO\textsubscript2(OH)\textsubscript2 & $Pnma$ & 1.49 & 3.64 & \\ \\
\hline
\end{tabular}
\raggedright
$^a$Ref.\citenum{Ohta:2002}. $^b$Ref.\citenum{Pellicer:2006}. $^c$Ref.\citenum{Rashad:2016}. $^d$Ref.\citenum{Diez:1995} (from reflectance peak at 470 nm). $^e$Ref.\citenum{Zhao:2015} (from absorbance edge at 710 nm). $^f$Ref.\citenum{Ozel:2016}. $^g$Ref.\citenum{Takayama:2017}. $^h$Ref.\citenum{Clarke:2008}. $^i$Ref.\citenum{Strobel:2008}. $^j$Ref.\citenum{Porter:2014}. $^k$Ref.\citenum{Porter:2015}.
\end{table}

In an effort to evaluate photocatalytic activity in connection to surface state formation, we developed a gas chromatography setup to measure hydrogen photogeneration. In analyzing these measurements, it should be borne in mind that the oxygen evolution reaction is much more sluggish than the hydrogen reduction reaction, and often requires a cocatalyst to proceed.\cite{Yang:2013} Although understanding the influence of cocatalysts on the photoactivity is of practical interest for optimizing solar-to-hydrogen conversion, this objective is beyond the scope of the present assessment whose goal is to examine the accuracy of intrinsic semiconductor properties within data-driven computational protocols. We thus restricted this analysis to the hydrogen reduction half-reaction by introducing sacrificial redox couples to circumvent the slow kinetics of oxygen evolution. The main outcome sought in these gas chromatography tests is to confirm the location of the conduction bands obtained from the Mott--Schottky measurements (\textit{i.e.}~$E_{\rm CB} < 0$ V vs.~RHE) and to assess the extent to which surface states may suppress photocatalytic activity.

Each gas chromatography test was performed by placing 10 mg of the synthesized crystalline powder into 5 mL of solution and exposing the system to illumination from a mercury arc lamp, providing light across the visible spectrum, with a fraction of ultraviolet contribution to also probe the wide-band optical response of some of the proposed materials. Gas concentrations were measured \textit{via} a valve-controlled gas chromatography setup, as described in Sec.~S6, ESI$^\dag$.  We examined two types of conditions: (\textit{i}) acidic pH, with the addition of 0.1 M of oxalic acid, which tends to favor the generation of H\textsubscript2 (by increasing the activity of the protons) but may also cause the premature dissolution of the sample; (\textit{ii}) neutral pH, corresponding to volume fractions of 15\% of methanol and 85\% of water. The magnitude of the H\textsubscript2 peak is then measured over time. The results of these series of experiments are presented in Table \ref{table:compilation} with the full gas chromatography responses given in Fig.~S6, ESI$^\dag$. A systematically assessment of the electrochemical stability of the tested compounds was carried out by means of XRD measurements and Pourbaix analysis. The results of this comprehensive assessment are reported in Sec. S8, ESI$^\dag$ and are discussed below.

A first observation is that although both of the identified plumbates exhibited some H\textsubscript2 signal; however, the generation of H$_2$ in both cases is accompanied with electrochemical corrosion. Interestingly, while the photoactivity of Ba\textsubscript2PbO\textsubscript4 decreases gradually due to electrodissolution, the rate of H\textsubscript2 production for Ca\textsubscript2PbO\textsubscript4 did not show any appreciable reduction, within the estimated experimental accuracy, over $\sim$100 hours of testing under condition (\textit{ii}). While Ca\textsubscript2PbO\textsubscript4 was previously identified theoretically,\cite{Castelli:2015} our study may be the first experimental validation of its photocatalytic proton-reduction activity and its favorable band alignment for water splitting.

Regarding the indate series, we observed that NaInO\textsubscript2 was photoactive, yielding hydrogen under  conditions (\textit{i}) and (\textit{ii}), and confirming previous literature evidence for wide-bang-gap hydrogen photocatalysis.\cite{Wang:2004} SrIn\textsubscript2O\textsubscript4 also showed a significant H\textsubscript2 signal under condition (\textit{i}), making it a potential lower-band-gap alternative to NaInO\textsubscript2. In contrast, CaIn\textsubscript2O\textsubscript4 did not exhibit any detectable photocatalytic activity, which provides further support to the possible formation of surface states that may promote charge recombination and affect interfacial charge transfer. In the same vein, we did not obtain a significant H\textsubscript2 response for BaIn\textsubscript2O\textsubscript4 under condition (\textit{ii}) and observed that the production of hydrogen under acidic condition (\textit{i}) was accompanied by surface corrosion (causing a change in the color of the sample). Upon surveying the literature, we found previous experimental confirmation of the photoactivity of SrIn\textsubscript2O\textsubscript4\cite{Sato:2001} for water splitting. The outcome of our gas-chromatograhy tests for NaInO\textsubscript2 and SrIn\textsubscript2O\textsubscript4 are nevertheless opposite to those reported in Ref.~\citenum{Sato:2001}, indicating that the mode of preparation and the potential occurrence of surface defects may be essential to the H\textsubscript2 photoactivity of this family of materials (in particular, the solid-state synthesis of CaIn\textsubscript2O\textsubscript4 may introduce surface states, as suggested by the Mott--Schottky measurements).  Likewise, we found in the literature that BaIn\textsubscript2O\textsubscript4 could promote photocatalysis when loaded with a RuO\textsubscript2 cocatalyst.\cite{Sato:2001} This result suggests that while some of the screened compounds were not confirmed to be photoactive (especially, due to corrosion in acidic environments), they may still be viable photocatalysts once combined with an auxiliary cocatalyst.

For completeness, we carried out the same battery of tests for the four materials containing open-shell Fe and Mn cations, finding in general a significant cathodic shifts in the measured flatband potentials relative to theoretical predictions. This systematic trend is consistent with possible Fermi-level pinning by mid-gap electronic states. While this shift precludes photocatalytic oxygen evolution for these Fe and Mn oxides, it brings the flatband potential of ZnFe\textsubscript2O\textsubscript4 and Na\textsubscript3Fe\textsubscript5O\textsubscript9 in close alignment with the redox potential of the H\textsubscript2/H$^+$ couple, yielding a large H\textsubscript2 signal in gas chromatography measurements. While the photocatalytic activity of ZnFe\textsubscript2O\textsubscript4 has recently been investigated and optimized,\cite{Kim:2015} our literature search did not reveal previous computational or experimental evidence of the photocathodic activity of Na\textsubscript3Fe\textsubscript5O\textsubscript9. As for BaCaFe\textsubscript4O\textsubscript8 and Ba\textsubscript3MnNb\textsubscript2O\textsubscript9, the shift in the redox potential appears to be too pronounced to enable proton reduction. 

To sum up, experimental measurements indicate favorable redox alignment and steady hydrogen generation for 6 of the 11 synthesized compounds, belonging to the plumbate and indate families. The absence of hydrogen production is limited to compounds exhibiting surface states or susceptible to photocorrosion. While this joint computational and experimental study provides a conclusive validation of the performance of the DFT+\textit U method without empirical fitting, it also calls for caution in considering magnetic order, defect levels, and surface passivation, especially in materials with open-shell transition metals.

\subsection{\label{section:refinement} Materials recommendations}

We now close the data-driven screening cycle by refining the search criteria to make final materials recommendations. To this end, we examined the 71 materials initially screened (Fig.~\ref{figure:protocol}) with the exception of the 11 compounds that we already synthesized and tested. We then narrowed down this list by restricting the candidates to closed-shell ions (\textit{d}\textsuperscript{0} or \textit{d}\textsuperscript{10}). We also did not retain halides due to their relatively poor aqueous stability compared with, \textit{e.g.}~oxides, nitrides or phosphides.~\cite{Su:2017} We thus obtained 4 binaries, 13 ternaries, and 5 quaternaries. 

By examining these materials, it can be first noted that all of the proposed binaries, namely, GaN, PbO, MnO, In\textsubscript2O\textsubscript3, were previously used for water photoelectrolysis or photocatalytic hydrogen reduction,~\cite{Chen2017,Osterloh2008,Gurudayal2016,Gyoung2018,Wang2005} which suggests that the refined search criteria are reliable to identify photocatalytic semiconductors.  Focusing next on the recommended ternaries and quaternaries, which are listed in Table~\ref{table:recommendations}, we observed that several of the identified materials are oxycuprates with a tendency to form layered structures due to the low coordination of their cuprous (Cu\textsuperscript{+}) ions. These cuprous oxide compounds include SrCu\textsubscript2O\textsubscript2 and BaCu\textsubscript2O\textsubscript2, and CuAlO\textsubscript2 and CuGaO\textsubscript2 (which adopt the 3$R$-type delafossite structure, $R\bar 3m$). The DFT+\textit U band gaps of these compounds are in agreement with experimental data (and consistent with the optical measurements presented in Fig.~\ref{figure:bandgap}), at variance with DFT values, which are significantly underestimated. Other oxides of interest similarly combine \textit{p}-block elements and group I-II metals, namely, mayenite 6CaO\textsubscript2-7Al\textsubscript2O\textsubscript3  (Ca\textsubscript{12}Al\textsubscript{14}O\textsubscript{33}), Na\textsubscript3BiO\textsubscript4, and Sr\textsubscript2PbO\textsubscript4.  

Beyond these oxides, the refined search enables us to identify 4 cuprous sulfides, one of them containing \textit d\textsuperscript{10} Sb\textsuperscript{3+} ions (Cu\textsubscript3SbS\textsubscript3) and the other three involving a series of \textit d\textsuperscript{0} early transition-metal ions W\textsuperscript{6+}, Nb\textsuperscript{5+}, and Y\textsuperscript{3+} (Cu\textsubscript2WS\textsubscript4, Cu\textsubscript3NbS\textsubscript4, CuYS\textsubscript2). In terms of crystal structure, it is worth noting that the latter compounds exhibit a gradual transition from a layered coordination (Cu\textsubscript2WS\textsubscript4) to a partially interconnected (sulvanite) structure (Cu\textsubscript3NbS\textsubscript4) to a three-dimensional covalent geometry (CuYS\textsubscript2), which are expected to influence their electronic bands. Despite these notable structural changes, all of these materials exhibit a covalently connected cuprous backbone, which may be at the origins of their narrow band gaps \textit{via} the formation of hybridized electronic states of Cu-3\textit{d} and S-3\textit{p} character near the valence band maximum.\cite{Takata:2017} This trend is captured by DFT+\textit U calculations within half of an eV, suggesting that electron localization on the sulfur site may play a critical role for sulfides, notwithstanding significant improvement over DFT predictions (which instead underestimate the band gap by more than 1 eV). In addition, the refined criteria enable us to identify oxychalgonide and oxynitride compounds that feature \textit d\textsuperscript{0} La\textsuperscript{3+} ions within covalently bonded layers. Of particular note are the oxynitrides LaTiO\textsubscript2N and CaTaO\textsubscript2N, which both exhibit a narrow band gap and have been shown to split water under visible light,~\cite{Chen2017,Minegishi2013,Xu2015} providing further confirmation of the efficacy of the screening approach. 

By carrying out a systematic literature search for the 18 recommended compounds, we found that 7 of them, namely, SrCu\textsubscript2O\textsubscript2,\cite{Wang:2011} CuGaO\textsubscript2,\cite{Zhao:2020} Sr\textsubscript2PbO\textsubscript4,\cite{Zhao:2015}Cu\textsubscript3SbS\textsubscript3,\cite{Ghorpade:2018} Cu\textsubscript3NbS\textsubscript4,\cite{Ibeka:2019} CaTaO\textsubscript2N,\cite{Xu:2015} LaTiO\textsubscript2N,\cite{Matsukawa:2014} have been experimentally identified as water-splitting photocatalysts, while CuAlO\textsubscript2\cite{Koriche:2005} and Cu\textsubscript2WS\textsubscript4\cite{Patir:2019} has been shown to promote photocatalytic hydrogen evolution. Moreover, Ca\textsubscript{12}Al\textsubscript{14}O\textsubscript{33}\cite{Liu:2018} and BaCu\textsubscript2O\textsubscript2\cite{Halasz:1990} are known photocatalysts for the reduction of methylene blue and the oxidation of carbon monoxide, respectively. To the extend of our literature search, none of the other 7 candidates (Na\textsubscript3BiO\textsubscript4, CuYS\textsubscript2, LaOCuS, LaOCuSe, La\textsubscript4O\textsubscript4Se\textsubscript3,
Na\textsubscript2TeO\textsubscript4, and Na\textsubscript5CuO\textsubscript2(OH)\textsubscript2) have yet been tested experimentally for photocatalytic water splitting.

\section{\label{section:conclusion}Conclusions}

We presented a comprehensive assessment of the reliability of data-driven materials screening for the discovery of water-splitting photocatalysts by comparing DFT+\textit U predictions (where the \textit U parameters were calculated using a fully automated, nonempirical linear-response method) to sensitive experimental measurements for 11 compounds out of an initial list of 70,150 candidates. These compounds were characterized by Mott--Schottky analysis and tested by gas chromatography, with 6 of them exhibiting steady-rate photocatalytic hydrogen evolution. Our computational and experimental results suggested that Ca\textsubscript2PbO\textsubscript4 could be catalytic for overall water splitting, and that both Ba\textsubscript2PbO\textsubscript4 and Na\textsubscript3Fe\textsubscript5O\textsubscript9 could be efficient photocathodes in promoting the photocatalytic reduction of hydrogen. To the extent of our literature search, these three materials have so far received limited attention as potential photocatalysts. Further electrocorrosion analysis revealed that Ba\textsubscript2PbO\textsubscript4 undergoes some initial electrochemical dissolution, which appears to preserve and possibly enhance its photocatalytic activity. We also found that Na\textsubscript3Fe\textsubscript5O\textsubscript9 is stable in aqueous solution, while Ba\textsubscript2PbO\textsubscript4 is likely to be unstable in humid atmosphere and aqueous solution.

At the computational level, our results point out that the reliability of electronic-structure predictions is critically dependent on the accurate description of the magnetic structure for open-shell transition-metal compounds. Beyond the importance of magnetic order, an additional level of complexity for this class of materials is their tendency to host mid-gap defect states that affect optical absorption and carrier lifetimes. At the practical level, our study highlights the primary relevance of \textit d\textsuperscript{0} and \textit d\textsuperscript{10} cations combined with alkali and alkaline-earth elements to develop water-splitting photocatalysts, and demonstrates the predictive performance of the proposed DFT+\textit U for this promising class of compounds. 

Additionally, using refined screening criteria based on our validation experiments, we recommended 18 materials, which include cuprous oxides, sulfides, and oxychalcogenides. Among these candidates, 7 compounds (Na\textsubscript3BiO\textsubscript4, CuYS\textsubscript2, LaOCuS, LaOCuSe, La\textsubscript4O\textsubscript4Se\textsubscript3,
Na\textsubscript2TeO\textsubscript4, Na\textsubscript5CuO\textsubscript2(OH)\textsubscript2) appear to not have been extensively studied as water-splitting photocatalysts and may deserve further theoretical and experimental consideration.

\section{\label{sec:methods}Methods}

\subsection{Electronic-structure calculations}

Electronic-structure calculations were performed using \textsc{Quantum ESPRESSO}.~\cite{Giannozzi2009, Giannozzi2017, Giannozzi2020} We employed the generalized gradient approximation (GGA) for the exchange-correlation functional with the Perdew--Burke--Ernzerhof (PBE) parameterization,~\cite{Perdew:1996} and ultrasoft pseudopotentials from the GBRV library.~\cite{Garrity2014} Kinetic energy cutoffs of 90~Ry for the wave functions and 720~Ry for the charge density and potentials were used. The Brillouin zone was sampled using a  Monkhorst--Pack~\cite{Monkhorst1976} $\bf k$-point mesh with a spacing of 0.04 \AA$^{-1}$. For all structures, atomic positions and lattice parameters were fully optimized at the GGA level, before any Hubbard correction. We performed DFT+\textit{U} calculations using the simplified formulation of Dudarev and coworkers.~\cite{Dudarev1998} The Hubbard correction~\cite{Anisimov1991, Anisimov1997, Dudarev1998} was applied to the \textit{d} (or $f$) states of transition-metal and rare-earth elements, and to the \textit{p} orbitals of oxygen and nitrogen. Within linear-response theory, the Hubbard parameters are the elements of an effective interaction matrix, evaluated as the difference between the bare and screened inverse susceptibilities:~\cite{Cococcioni2005} 
\begin{equation}
U^I = \left(\chi_0^{-1} - \chi^{-1}\right)_{II} \,,
\label{eq:Ucalc}
\end{equation}
where $I$ is the atomic site index. The susceptibilities $\chi_0$ and $\chi$ were computed from the response of atomic occupations to shifts in the potential acting (through projectors) simultaneously on the relevant orbitals of the isolated atom: $\chi_{IJ} = \sum_{m\sigma} \left(dn^{I\sigma}_{mm} / d\alpha^J\right)$, where $n^{I\sigma}_{mm'}$ are atomic occupation matrices, $\alpha^J$ is the strength of the perturbation on the $J$\textsuperscript{th} site, $m$ and $m'$ are magnetic quantum numbers associated with a specific angular momentum, and $\sigma$ is the spin index. The response $\chi$ is evaluated at self-consistency (of the linear-response Kohn-Sham calculation), while $\chi_0$ is computed before the self-consistent re-adjustment of the Hartree and exchange-correlation potentials. Using DFPT, the response to isolated perturbations can be evaluated as the sum of monochromatic ($\mathbf{q}$-specific) contributions, computed independently on a grid of $\mathbf{q}$ points of the Brillouin zone, from calculations on the primitive unit cell~\cite{Timrov2018}:
\begin{equation}
\frac{dn^{I\sigma}_{mm'}}{d\alpha^J} = \frac{1}{N_{\mathbf{q}}}\sum_{\mathbf{q}} e^{i\mathbf{q}\cdot(\mathbf{R}_l - \mathbf{R}_{l'})}\Delta_{\mathbf{q}}^{s'} n^{s \,\sigma}_{mm'} .
\label{eq:dnq}
\end{equation}
In this equation, $I\equiv(l,s)$ and $J\equiv(l',s')$, $l$ and $l'$ label the unit cells, $s$ and $s'$ label atoms in the unit cells, $\mathbf{R}_l$ and $\mathbf{R}_{l'}$ are Bravais lattice vectors, and $\Delta_{\mathbf{q}}^{s'} n^{s \,\sigma}_{mm'}$ represent the lattice-periodic response of atomic occupations to monochromatic perturbations constructed by modulating the shift to the potential of all the periodic replica of a given atom by a wavevector $\mathbf{q}$. The quantities $\Delta_{\mathbf{q}}^{s'} n^{s\sigma}_{mm'}$ were obtained by solving DFPT equations, independently for every $\mathbf{q}$.~\cite{Timrov2018} In periodic systems, this approach allows to eliminate the need for supercells for computing \textit{U}.~\cite{Timrov2018} The calculations of the \textit{U} parameters using DFPT were performed with a single $\mathbf{q}$-point. We ran convergence tests with denser $\mathbf{q}$-point meshes of 2$\times$2$\times$2 and 4$\times$4$\times$4, and found the band gap value only changed by $\sim$0.1~eV, an acceptable loss in accuracy for the time savings in a high-throughput workflow. To construct the projectors of the DFT+\textit{U} scheme, we used atomic orbitals that were orthogonalized by applying the L\"owdin method.~\cite{Lowdin1950, Mayer2002}

\subsection{Materials synthesis}

All samples were synthesized by finely grinding and pelletizing a mixture of powders using an agate mortar and pestle in the molar ratios described below. The samples were added to an alumina boat and heated in air either in a Mullite tube furnace or a Lindberg/Blue M tube furnace, as indicated for each sample below. The samples were heated at 5 $^\circ$C/min and held at 400 $^\circ$C and 800 $^\circ$C for two hours prior to heating to the final temperature indicated for each sample, unless other parameters are explicitly mentioned. The samples were then cooled to room temperature inside the furnace.

\textit{Synthesis of Ca\textsubscript2PbO\textsubscript4 powder}:~CaCO\textsubscript3 powder (Alfa Aesar, 99.99\%) and PbO powder (Alfa Aesar, 99.999\%) were combined in a 2:1 molar ratio of CaCO\textsubscript3:PbO and heated to 800 $^\circ$C for 26 hours in a Lindberg/Blue M tube furnace. Note that the PbO used to produce Ca\textsubscript2PbO\textsubscript4 had an orange color, likely due to Pb\textsubscript2O\textsubscript3 impurities; Pb\textsubscript2O\textsubscript3 was necessary for this phase to form in high yield.
\textit{Synthesis of Ba\textsubscript2PbO\textsubscript4 powder}:~BaCO\textsubscript3 powder (Alfa Aesar, 99.95\%) and PbO powder (Alfa Aesar, 99.999\%) were combined in a 2:1 molar ratio and heated to 1,100 $^\circ$C for 24 hours in a Lindberg/Blue M tube furnace.
\textit{Synthesis of NaInO\textsubscript2 powder}:~Na\textsubscript2CO\textsubscript3 powder (EMD Chemicals, 99.9\%) and In\textsubscript2O\textsubscript3 powder (Alfa Aesar, 99.99\%) were combined in a 1:1 molar ratio and heated to \linebreak 900 $^\circ$C for 3 hours in a Lindberg/Blue M tube furnace.
\textit{Synthesis of CaIn\textsubscript2O\textsubscript4 powder}:~CaCO\textsubscript3 powder (Alfa Aesar, 99.99\%) and In\textsubscript2O\textsubscript3 powder (Alfa Aesar, 99.99\%) were combined in a 1:1 molar ratio and heated to 1,050 $^\circ$C for 12 hours in a Lindberg/Blue M tube furnace.
\textit{Synthesis of SrIn\textsubscript2O\textsubscript4 powder}:~SrCO\textsubscript3 powder (Alfa Aesar, 99.99\%) and In\textsubscript2O\textsubscript3 powder (Alfa Aesar, 99.99\%) were combined in a 1:1 molar ratio and heated to 1,050 $^\circ$C for 12 hours in a Lindberg/Blue M tube furnace.
\textit{Synthesis of BaIn\textsubscript2O\textsubscript4 powder}:~BaCO\textsubscript3 powder (Alfa Aesar, 99.95\%) and In\textsubscript2O\textsubscript3 powder (Alfa Aesar, 99.99\%) were combined in a 1:1 molar ratio and heated to 1,050 $^\circ$C for 12 hours in a Lindberg/Blue M tube furnace.
\textit{Synthesis of PbTiO\textsubscript3 powder}:~PbO powder (Alfa Aesar, 99.999\%) and TiO\textsubscript2 powder (Alfa Aesar, 99.9\%) were combined in a 1:1 molar ratio and heated to 900 $^\circ$C for 12 hours in a Lindberg/Blue M tube furnace.
\textit{Synthesis of ZnFe\textsubscript2O\textsubscript4 powder}:~ZnO powder (Sigma Aldrich, $\ge$99.0\%) and Fe\textsubscript2O\textsubscript3 powder (Aldrich, catalyst grade) were combined in a 1:1 molar ratio and heated to 900 $^\circ$C for 72 hours in a Lindberg/Blue M tube furnace.
\textit{Synthesis of Na\textsubscript3Fe\textsubscript5O\textsubscript9 powder}:~Na\textsubscript2CO\textsubscript3 powder (EMD Chemicals, 99.9\%) and Fe\textsubscript2O\textsubscript3 powder (Aldrich, catalyst grade) were combined in a 3:5 molar ratio and heated to 1,100 $^\circ$C for 48 hours in a Mullite tube furnace.
\textit{Synthesis of BaCaFe\textsubscript4O\textsubscript8 powder}:~Ba(NO\textsubscript3)\textsubscript2 powder (Sigma-Aldrich, $\ge$99\%), CaCO\textsubscript3 powder (Alfa Aesar, 99.99\%), and FeO (Alfa Aesar, 99.95\%) were combined in a 1:1:4 molar ratio and heated to 1100 $^\circ$C for 48 hours in a Lindberg/Blue M tube furnace. The sample was cooled to room temperature, reground, pelletized, and heated to 1,100  $^\circ$C for an additional 48 hours.
\textit{Synthesis of Ba\textsubscript3MnNb\textsubscript2O\textsubscript9 powder}:~BaCO\textsubscript3 powder (Alfa Aesar, 99.95\%), MnO\textsubscript2 powder (Sigma-Aldrich, $\ge$99\%), and Nb\textsubscript2O\textsubscript5 powder (Sigma-Aldrich, 99.99\%) were combined in a 6:2:1 molar ratio and heated to 1,300 $^\circ$C for two days in a Mullite tube furnace.

\subsection{Materials characterization}

\textit{X-ray diffraction}:~Powder X-ray diffraction (XRD) was performed on a Malvern PANalytical Empyrean (3rd gen.) X-ray Diffractometer for 2$\theta$ in the range of 20$^\circ$ to 80$^\circ$. The pellets of each material were ground to powders prior to analysis. Reference XRD patterns were generated from either the Powder Diffraction File (PDF) card number or crystallographic data:
Ca\textsubscript2PbO\textsubscript4: PDF 04-008-2917;
Ba\textsubscript2PbO\textsubscript4: PDF 04-007-5957;
NaInO\textsubscript2: PDF 04-008-3834;
CaIn\textsubscript2O\textsubscript4: The CaIn\textsubscript2O\textsubscript4 structure was constructed by substituting Ca into the SrIn\textsubscript2O\textsubscript4 structure (PDF 04-013-8519) and adjusting the volume of the cell to match the experimental data using cell parameters (space group $Pnma$) of $a$ = 9.68 \AA, $b$ = 11.30 \AA, and $c$ = 3.22 \AA, with $\alpha$ = $\beta$ = $\gamma$ = 90 $^\circ$C and a cell volume of 352.2 \AA\textsuperscript{3};
SrIn\textsubscript2O\textsubscript4: PDF 04-013-8519;
BaIn\textsubscript2O\textsubscript4: PDF 04-013-8196;
PbTiO\textsubscript3: PDF 04-006-5418;
ZnFe\textsubscript2O\textsubscript4: PDF 04-002-2708;
Na\textsubscript3Fe\textsubscript5O\textsubscript9: PDF 04-011-2582;
BaCaFe\textsubscript4O\textsubscript8: PDF 00-018-0147;
Ba\textsubscript3MnNb\textsubscript2O\textsubscript9: PDF 00-046-0998

\textit{Diffuse reflectance}:~The samples were ground in a mortar with ethanol, and then drops of this suspension were placed on glass slides and left to dry. Consecutive drops were added until a uniform, thick, and opaque film of the powders was observed (based on lack of light transmission through the film). A Perkin Elmer lambda 950 was employed to measure diffuse reflectance spectra using a 150 mm integrating sphere collecting data from 250-2500 nm, taking 1-nm steps, and using a 4-nm slit width in diffuse reflection mode. The reference spectrum for total reflectance was measured against a Spectralon disc. A plot of the Kubelka--Munk function, raised to the power of $\frac 12$ or 2 for indirect and direct semiconductors, respectively, as a function of energy (in nm) was constructed to obtain the band gaps. These band gaps were calculated using the derivative of the Kubelka--Munk plot, finding the linear region at the onset of absorption from high to low energy, and extrapolating the region to the intercept along the energy axis.

\textit{Electrode preparation}:~The pellets were crushed into powders that were subsequently ball-milled (using high-density zirconium oxide balls) to improve their dispersion in an ethanol suspension for 24 hours. Ethanol-powder inks normalized to 0.002 mmol/mL were deposited on 5$\times$8$\times$1.1-mm thick TEC7 Fluorine-doped tin oxide (FTO) conductive substrates. Two batches of electrodes were made, one with 100 $\mu$L of ink deposited and the other with 120 $\mu$L of ink deposited. The slides were then annealed at 400 $^\circ$C for two hours. To construct the working electrodes, the FTO slides were placed on regular glass slides and ohmic contacts were made using silver paint between the slide and a piece of copper tape. The electrodes were then insulated and secured using epoxy.

\textit{Mott--Schottky measurements}:~Measurements were carried out on a Biologic SP-300 potentiostat using the `Staircase Potentio Electrochemical Impedance Spectroscopy' feature in a pH-8 aqueous sodium phosphate buffer. These measurements were done over a range of potentials at constant frequency.  The typical analyses were obtained at 20,000 Hz, 16,666 Hz, 13,333 Hz, and 9,999 Hz with a sinus amplitude frequency of 7 mV. The voltage sweep range was chosen to be within 0.5 to 1.5 V of the expected flatband potential based on previous open circuit values. Both 100 $\mu$L and 120 $\mu$L dropcast film electrodes were tested in open-circuit conditions and under illumination.

\textit{Gas chromatography}:~The hydrogen reduction reaction analysis was carried out using a self-built setup that contains a reaction chamber and a gas chromatograph. The setup is depicted in Sec.~S6, ESI$^\dag$. In each test, 10 mg of the synthesized powder was dispersed in 5 mL of aqueous solution (\textit{i}) with the addition of 0.1 M of oxalic acid and (\textit{ii}) with a volume fraction of 15\% of methanol, under Argon flow with a partial pressure of 1 atm. The sample was finally illuminated with a 200 W arc lamp from ORIEL with a wavelength of 200-800 nm for a period of time. An 800 nm cutoff filter was applied to avoid heating. The generated gas was then pumped to the gas chromatograph HP 5890 series II using thermal conductivity detector under argon carrier gas. The results for each tested photocatalyst are shown in Fig.~S5, ESI$^\dag$.

\section*{Author contributions}

Y.~X.~and Q.~T.~C.~contributed equally, as first authors. Y.~X., Q.~T.~C., and N.~E.~K.~derived the criteria of the computational screening and implemented the software infrastructure. J.~S.~M.~initiated the synthesizability analysis, which was finalized by J.~F.~and consolidated into a database by N.~E.~K.. Q.~T.~C., N.~C.~S., and X.~Q.~performed the preliminary high-throughput calculations, which were finalized, validated, and curated by Y.~X., N.~E.~K., M.~M.~K., B.~P., and P.~O.. I.~T.~and M.~C.~developed and implemented the \textit{ab initio} Hubbard \textit U method, and ported it into the computational workflow. J.~S.~M., and K.~S.~carried out the initial synthesis and XRD analysis. J.~F.~optimized the phase uniformity of the synthesized materials, and characterized their structural and optical properties. R.~K.~and Y.~X.~carried out the comparative experimental and computational analysis of the electrochemical stability of the compounds.  H.~W.~developed the gas chromatography setup and tested the photoactivity of the compounds. C.~K.~B.~and A.~M.~V.~performed the preliminary photoelectrochemical experiments. C.~K.~B.~carried out the Mott--Schottky experiments at the National Renewable Energy Laboratory, which were then finalized by M.~J.~T.~and C.~K.~B.~at Cornell University. N.~E.~K., Y.~X., and Q.~T.~C.~surveyed the high-throughput computational literature and previous studies on the recommended materials. S.~A.-H.~and N.~E.~K.~researched and developed quantitative criteria for the toxicology analysis, and implemented them in the screening protocol. C.~J.~F., S.~A.-H., J.~L.~Y., T.~G.~D., M.~C., V.~G., H.~D.~A., R.~E.~S., and I.~D.~supervised the work. Y.~X., Q.~T.~C., J.~F., C.~K.~B., H.~W., N.~E.~K., I.~T., A.~M.~V., S.~A.-H., M.~C., V.~G., H.~D.~A., R.~E.~S., and I.~D.~drafted the manuscript with contributions from each author. The manuscript was thoroughly revised by all the authors at each stage of its preparation.

\section*{Conflicts of interest}

The authors declare no competing financial interest.

\section*{Acknowledgements}

This work was primarily supported by the DMREF and INFEWS programs of the National Science Foundation under Grant No.~DMREF-1729338. I.~T. acknowledges support from the Swiss National Science Foundation, through Grant No.~200021-179138, and its National Centre of Competence in Research (NCCR) MARVEL. I.~T.~and M.~C.~also acknowledge partial support from the EU-H2020 Research and Innovation Programme under Grant Agreement No.~654360 within the framework of the NFFA Europe Transnational Access Activity. B.P.~and C.G.~acknowledge partial support from the NSF Platform for the Accelerated Realization, Analysis, and Discovery of Interface Materials (PARADIM) under Cooperative Agreement No.~DMR-1539918. I.~D.~acknowledges partial support for computational resources from the Corning Faculty Fellowship in Materials Science and Engineering. This work was authored in part by the National Renewable Energy Laboratory (NREL), operated by Alliance for Sustainable Energy, LLC, for the U.S. Department of Energy (DOE) under Contract No.~DE-AC36-08GO28308. The views expressed in the article do not necessarily represent the views of the DOE or the U.S. Government. The U.S. Government and the publisher, by accepting the article for publication, acknowledge that the U.S. Government retains a nonexclusive, paid-up, irrevocable, worldwide license to publish or reproduce the published form of this work, or allow others to do so, for U.S. Government purposes. The authors acknowledge research support from the HydroGEN Advanced Water Splitting Materials Consortium, established as part of the Energy Materials Network under the U.S. Department of Energy, Office of Energy Efficiency and Renewable Energy, Fuel Cell Technologies Office, under Contract No.~DE-AC36-8GO28308 to the NREL. This work was supported in part by the U.S. Department of Energy, Office of Science, Office of Workforce Development for Teachers and Scientists (WDTS) under the Science Undergraduate Laboratory Internships Program (SULI), and by the National Science Foundation through the Research Experiences for Undergraduates and Research Experiences for Teachers in Nanoscale Physics and Materials at Pennsylvania State University under Grant No.~DMR-1851987. The authors are deeply thankful to J.~H.~Golbeck, T~.E.~Mallouk, and N.~Marzari for fruitful discussions.

\section*{Data availability}

The data generated in this study are reported in the main text and in the ESI$^\dag$, with references provided for all data that were obtained from the literature. Additional data and metadata are available from the authors upon request.  



\balance




\providecommand*{\mcitethebibliography}{\thebibliography}
\csname @ifundefined\endcsname{endmcitethebibliography}
{\let\endmcitethebibliography\endthebibliography}{}

\end{document}